# Inhibited nonradiative decay at all exciton densities in monolayer semiconductors

*Hyungjin Kim[1,2,†], Shiekh Zia Uddin[1,2,†], Naoki Higashitarumizu[1,2], Eran Rabani[2,3,4] and Ali Javey[1,2,*]*

[1]Electrical Engineering and Computer Sciences, University of California, Berkeley, CA 94720, USA
[2]Materials Sciences Division, Lawrence Berkeley National Laboratory, Berkeley, CA 94720, USA
[3]Department of Chemistry, University of California, Berkeley, Berkeley, California 94720, USA
[4]The Raymond and Beverly Sackler Center of Computational Molecular and Materials Science, Tel Aviv University, Tel Aviv 69978, Israel

[†]These authors contributed equally
*Address correspondence to: ajavey@berkeley.edu

**Most optoelectronic devices operate at high photocarrier densities, where all semiconductors suffer from enhanced nonradiative recombination. Nonradiative processes proportionately reduce photoluminescence (PL) quantum yield (QY), a performance metric that directly dictates the maximum device efficiency. Although transition-metal dichalcogenide (TMDC) monolayers exhibit near-unity PL QY at low exciton densities, nonradiative exciton-exciton annihilation (EEA) enhanced by van-Hove singularity (VHS) rapidly degrades their PL QY at high exciton densities and limits their utility in practical applications. Here, by applying small mechanical strain (< 1%), we circumvent VHS resonance and drastically suppress EEA in monolayer TMDCs, resulting in near-unity PL QY at all exciton densities despite the presence of a high native defect density. Our findings can enable light-emitting devices that retain high efficiency at all brightnesses.**

Exciton recombination in TMDC monolayers depends on exciton generation rate, background carrier concentration, and electronic band structure. The combined effect of the first two factors has been studied thoroughly *(1)*, where the photocarrier generation rate (*G*) and background carrier concentration were tuned by varying the incident pump power and the gate voltage ($V_g$), respectively, in a capacitor structure. Given pronounced Coulomb interactions, background carriers turn photogenerated excitons into charged trions *(2)* that mostly recombine nonradiatively *(1, 3)*. In the absence of background carriers, at low exciton densities, neutral excitons in intrinsic monolayers can recombine completely radiatively even in the presence of defects *(1, 4)*. However, at high exciton densities, recombination of neutral excitons is dominated by exciton−exciton annihilation (EEA), where an exciton nonradiatively recombines while colliding with another exciton *(4-6)*. All excitonic materials found in nature *(7)* exhibit EEA, which is the primary cause of efficiency roll-off observed in all organic *(8)* and some inorganic *(9,10)* light-emitting devices. The EEA effect has similarities to Auger recombination, which is commonly observed in conventional free-carrier systems and also a prominent cause of efficiency roll-off observed in light-emitting diodes and solar cells *(11, 12)*. Through the conservation of momentum and energy the participating quasiparticles, EEA depends on the aforementioned third factor: detailed band structure *(6, 13, 14)*.

In this work, we modulate these three factors simultaneously. Along with $V_g$ and pump power, we altered the electronic band structure by applying a uniaxial strain ($\epsilon$). With appropriate strain, all neutral excitons recombined radiatively even at high concentrations in monolayers of $WS_2$, $WSe_2$, and $MoS_2$, resulting in near-unity quantum yield (QY) at all measured generation rates. The density of states of electrons in a two-dimensional (2D) periodic crystal are topologically constrained to exhibit logarithmic van Hove singularities (VHSs) arising from saddle points in the energy dispersion *(15)*. When the energy of a transition is near the VHS, weak interactions are often intensified by the enhanced density of states (DOS) *(16)*. From energy and momentum conservation, we showed that as-exfoliated TMDC monolayers exhibited enhanced EEA because the final energy of this process coincided with inherent VHS. Strain drove the final energy away from the VHS resonance and drastically reduced EEA in both sulfur- and selenide based TMDCs. We could uniformly suppress all nonradiative recombination in a centimeter-scale CVD-grown $WS_2$ monolayer at all generation rates.

Monolayer field effect transistor (FET) devices were fabricated on plastic flexible substrate enabling simultaneous modulation of carrier concentration and strain while performing quantitative photoluminescence (PL) QY measurements (device fabrication details are available in the supplementary materials). Fig. 1A shows the schematic and optical micrograph of such a device using WS$_2$ monolayers. Mechanically exfoliated monolayers were transferred to a polyvinyl alcohol (PVA) layer attached to polyethylene terephthalate (PET) handling substrate, a combination selected for its superior strain transfer efficiency *(17)*. Exfoliated hexagonal boron nitride (hBN; 70 to 100 nm in thickness) and graphene (2 to 5 nm in thickness) were transferred sequentially on the monolayer WS$_2$ to serve as a gate insulator and gate electrode, respectively. The WS$_2$ monolayer was electrically grounded and $V_g$ was applied to the top graphene. Bending the PET substrate with positive curvature applied tensile strain to the monolayer in the direction of bending *(18)*. Applied strain is calculated from geometrical considerations. Generation rate $G$ is the number of excitons created or the number of photons absorbed per-unit area per unit time, which can be tuned by changing the laser intensity.

As-exfoliated monolayer WS$_2$ is known to be electron-rich. Application of a negative $V_g(-30\,V)$ electrostatically compensated for that unintentional electron doping and ensured that the recombination process was dominated by neutral excitons *(1)*. At this biasing condition, we compared PL spectra of an unstrained and strained WS$_2$ monolayer at a high generation rate of $G = 6.5 \times 10^{19}\,cm^{-2}s^{-1}$ (Fig. 1B). We observed both a redshift of PL spectra and ~15 times enhancement in PL intensity with the application of $\epsilon = 0.4\%$ tensile strain at this high generation rate. Note that, $G$ is independent of strain since absorption at pump wavelength does not change with strain (fig. S1).

Calibrated PL measurements at room temperature were performed to quantitatively extract the QY as a function of $\epsilon$, $V_g$ and $G$ (Fig. 1, C to F, fig. S2), and Fig. 1C shows PL QY as a function of the gate voltage, $V_g$ and the generation rate, $G$ when no strain is applied. At zero and positive gate voltages, background electron concentration of monolayer WS$_2$ was large and led to formation of negative trions that predominantly recombined nonradiatively, thus yielding a low PL QY. At negative $V_g$ the background electrons were removed, and recombination of neutral excitons dominated. At low generation rates, neutral excitons recombined completely radiatively despite native defect density.

At high generation rates, PL QY rolled off because of EEA. These same results were observed previously for other substrates, such as PMMA and SiO$_2$ and are quantitatively the same as the results here *(1)*. With the application of 0.2% tensile strain, the PL QY drop-off at high exciton generation rate was notably reduced for negative $V_g$. (Fig. 1D). For a tensile strain of 0.4%, no PL QY drop-off at high generation rate was observed (Fig. 1E). Thus, all nonradiative recombination processes in as-exfoliated monolayer WS$_2$ were suppressed by applying tensile strain and $V_g$.

Suppression of EEA is further elucidated in Fig. 1F, where we show PL QY as a function of strain at $V_g = -30\,V$ and at a high generation rate $G = 6.5 \times 10^{19}\,cm^{-2}s^{-1}$. For no applied strain, PL QY was low at this generation rate, but for a threshold strain of 0.3% PL QY increased sharply and asymptotically approached unity. This PL QY enhancement by strain was also reversible and repeatable, as PL QY traces for increasing and decreasing strain fully overlapped (fig. S3). Such reversibility and repeatability indicates absence of slippage, so applied strain is equal to the actual strain transferred to the monolayer. High PL QY at all pump powers persisted even for a tensile strain of $\epsilon = 1.0\%$ (Fig 1F, fig. S4). In addition to electrostatic counter-doping, high PL QY was achieved by applying tensile strain in monolayer $WS_2$ chemically counter-doped by Nafion, a known hole-dopant (fig. S5).

The PL QY for neutral excitons ($V_g = -30\,V$) can be written as the ratio of radiative recombination rate of to the total recombination rate

$$QY = \frac{R_r}{R_r + R_{nr}} \qquad (1)$$

where $R_r$ and $R_{nr}$ are the exciton radiative and nonradiative recombination rates, respectively. Nonradiative recombination for neutral excitons are predominantly through the EEA process *(1)*. Therefore, $R_{nr} = R_{EEA}$ where $R_{EEA}$ is the nonradiative EEA rate. Even with application of strain, the semiconductor remained strongly excitonic *(6)*, so strain did not change the PL QY versus $G$ response at low generation rates, and $R_r$ was independent of strain (Fig. 1, D to F. However, because strain changed the QY at high pump powers, $R_{EEA}$ must depend strongly on the strain. Using the experimentally measured PL QYs at a high generation rate of $G = 6.5 \times 10^{19}\,cm^{-2}s^{-1}$, we found that

$$\frac{R_{EEA}(\epsilon = 0.4\%)}{R_{EEA}(\epsilon = 0.0\%)} \approx 7 \times 10^{-3} \qquad (2)$$

which would indicate roughly two orders of magnitude decrease in EEA rate at the highest generation rate. Although the EEA rate was not rigorously zero, it was decreased by the application of strain such $R_r \gg R_{EEA}$, and the radiative relaxation dominated the total recombination.

The photophysics of these monolayer semiconductors as a function of $V_g$ and generation rate arises from the quasiparticle interaction and can be captured by a simple kinetic model described elsewhere *(1)*. The exciton-to-trion ratio can be tuned with $V_g$, activating their respective recombination routes, whereas high densities of neutral excitons activate the EEA process. However, the mechanism by which strain suppresses EEA at high pump necessitated a closer look at the EEA process itself. EEA occurs when one exciton ionizes another exciton by nonradiatively transferring its energy (Fig. 2A) *(6)*. The initial state consists of two excitons, with center-of-mass momenta $\boldsymbol{K_1}$ and $\boldsymbol{K_2}$ and energies $E_1$ and $E_2$, respectively. The exciton energy and momentum are related by *(13)*

$$E_1 = E_G - E_B + \frac{\hbar^2|\boldsymbol{K_1}|^2}{2M_X} \qquad (3)$$

$$E_2 = E_G - E_B + \frac{\hbar^2|\boldsymbol{K_2}|^2}{2M_X} \qquad (4).$$

In the above relations, $E_G$ is the fundamental bandgap, $E_B$ is the exciton binding energy and $M_X$ is the exciton mass. The final state consists of a high energy electron and hole, with crystal momenta $\boldsymbol{k_e}$ and $\boldsymbol{k_h}$ and energies $E_e$ and $E_h$, respectively.

Irrespective of the details of the interaction potential, two quantities are conserved in the EEA process: total momentum and energy *(13, 19, 20)*. The condition for conservation of crystal momentum yields *(13)*

$$k_e + k_h = K_1 + K_2. \quad (5)$$

Because $K_1$ and $K_2$ are determined by the thermal motion of excitons, they are negligible compared to $k_e$ and $k_h$, implying that the electron and the hole momentum in the final state should be almost opposite to each other.

$$k_e \approx -k_h, \quad (6)$$

Therefore, momentum conservation dictates the electron and hole from the ionized exciton have opposing crystal wavevectors (on top of each other in the energy dispersion, shown with the red dashed line in Fig. 2A). The condition for conservation of crystal momentum yields *(13)*

$$E_e + E_h = E_1 + E_2. \quad (7)$$

Because the exciton center of mass wavevector is negligible, $E_1 \approx E_2 \approx E_G - E_B = E_X$, where $E_X$ is the exciton transition energy. Therefore, energy conservation stipulates that the energy difference between the electron and hole must be $2E_X$

$$E_e + E_h \approx 2E_X \quad (8)$$

If we denote the conduction and valance band of the semiconductor as $E_C$ and $E_V$, then combined momentum and energy conservation can be written as

$$E_e + E_h = E_C(k_e) - E_V(-k_h) = E_C(k_e) - E_V(k_e) = 2E_X. \quad (9)$$

Therefore, any wavevector where the energy difference between the conduction and valance band is equal to twice the exciton transition energy can be the final wavevector of the electron from the ionized exciton (Fig. 2B). Note that the exciton transition energy is different from the fundamental bandgap because of enhanced electron-hole interaction in TMDC monolayers. By Fermi's golden rule, the joint density of states (JDOS) at twice the exciton transition energy determines the strength of EEA process *(13, 20)*.

Typically, the effective mass approximation is invoked to determine the energy of the ionized electron. However, the ionized electron and hole have high energy so this is no longer a valid approximation. We first calculated the band structure of monolayer TMDC with an 11-band tight binding theory based on Wannier transformation of ab-initio density functional theory calculations [details of band structure are in *(21, 22)*]. We then calculated the energy difference between the conduction and valance band ($E_C - E_V$) for unstrained and strained monolayer $WS_2$, respectively. The first Brillouin zone is indicated by a gray hexagon (Fig. 2, C and D). Red areas are the final states of the electron from the ionized exciton where the conservation laws are satisfied.

We note that possible final wavevectors in unstrained monolayer $WS_2$ include the saddle points Q (Fig. 2C), but in 0.4% strained $WS_2$ they do not include the saddle points (Fig. 2D). Saddle points in the band structure of a 2D semiconductor create VHS and result in a logarithmically diverging JDOS *(15, 23, 24)*. The JDOS for monolayer $WS_2$ $\rho_{CV}(E)$ at $E = E_C - E_V = 2E_X$ determines the strength of EEA. We show the JDOS of 0.4% strained and unstrained $WS_2$ in Fig 2E. In unstrained samples, there was a VHS at twice the exciton transition energy, resulting in an expedited EEA. Strain shifted the exciton transition energy $E_X$ such that $2E_X$ did not overlap VHS resonance and reduced EEA.

We also show the JDOS of a 0.4% compressively strained $WS_2$ monolayer in Fig. 2E. We note that compressive strain also drove the EEA process off VHS resonance by changing the exciton transition energy, both for uniaxial strain applied in any direction or biaxial strain (fig. S6, S7, S8). The choice of tensile versus compressive strain to suppress all nonradiative recombination should ensure that the system remains direct bandgap with the application of strain to avoid nonradiative recombination through momentum-dark indirect excitons *(25)*. Compressive strain can make some TMDC monolayers such as WS$_2$ indirect.

Strain has also been used to reduce traditional Auger recombination in conventional 3D semiconductors, but the mechanism is different *(26)*. In 3D semiconductors, bandgap renormalization and effective mass equalization by applied strain can lead to one order of magnitude reduction of conventional Auger rate *(27)*. However, in the case of 2D TMDCs EEA is inhibited by shifting the exciton transition energy $E_X$ such that $2E_X$ did not overlap VHS resonance. Because the existence of saddle points, logarithmic VHS always characteristically appeared in 2D semiconductors and were not found in three dimensions, leading to a much stronger response to strain in monolayer TMDCs *(15)*.

These principles applied equally to other TMDC semiconductors. Like WS$_2$, exfoliated monolayer WSe$_2$ also exhibited near-unity PL QY at all generation rates when it was made intrinsic by electrostatic counterdoping and tensile strain was being applied (Fig. 3A and fig S9). As-exfoliated monolayer MoS$_2$ has PL QY in the range of 0.1 to 1.0 %, which drastically increased at low pump after chemical counterdoping by Nafion (Fig. 3B). Unlike WS$_2$ and WSe$_2$, monolayer MoS$_2$ became an indirect-gap material when tensile strain was applied and remained direct-gap when compressive strain was applied *(17)*.

We found that, rather than creating compressive strain, downward bending of flexible substrate with negative curvature resulted in buckling of monolayer TMDC. Instead, we used the thermal coefficient of expansion mismatch between the glycol-modified polyethylene terephthalate (PETG) substrate and MoS$_2$ to apply compressive strain. A chemically counterdoped, 0.5% compressively strained monolayer MoS$_2$ also exhibited roll-off free PL QY at all generation rates.

The optoelectronic quality of large-area 2D TMDCs must be improved for their use as next-generation devices. The principles of suppressing all nonradiative recombination can also be applied to achieve high PL QY on centimeter-scale WS$_2$ monolayer grown by chemical vapor deposition (CVD). We first transfer a large-area CVD grown WS$_2$ *(28)* onto a flexible substrate and spin-coated it with Nafion (Fig. 4A). The normalized PL spectra redshifted with applied tensile strain (Fig. 4B). Nafion counterdoping led to a strain-independent PL QY of $70 \pm 10$ % (mean + standard deviation) at low pump powers, as can be seen in a spatial map of the PL QY of a 2 mm × 2 cm area taken at a generation rate of $G = 10^{16}\ cm^{-2}s^{-1}$ (Fig. 4C). At a high generation rate of $G = 10^{20}\ cm^{-2}s^{-1}$, if no strain was applied, a low PL QY of $1.6 \pm 0.8$ % was observed uniformly, consistent with as-exfoliated monolayer WS$_2$. (Fig. 4D). However, at 0.4% applied tensile strain, PL QY of the same area reaches $59 \pm 10$ %, indicating a ~38 times uniform enhancement of PL QY for a large-scale grown sample at a high generation rate of $G = 10^{20}\ cm^{-2}s^{-1}$. This demonstration showcases that ideal optoelectronic quality can be achieved in large-area grown monolayers by relatively simple and scalable application of counterdoping and strain.

**Funding**

This work is supported by the Director, Office of Science, Office of Basic Energy Sciences, Materials Sciences and Engineering Division of the U.S. Department of Energy, under contract no. DE-AC02-05Ch11231. H. K. acknowledges support from Samsung Scholarship. N. H. acknowledges support from the Postdoctoral Fellowships for Research Abroad of Japan Society for the Promotion of Science.

**Author Contributions**

S.Z.U., H.K., and A.J. conceived the idea for the project and designed the experiments. H.K. and S.Z.U. performed optical measurements. H.K., S.Z.U., and N.H. fabricated devices. H.K., S.Z.U., and A.J. analyzed the data. S.Z.U. and E.R. performed analytical modeling. S.Z.U., H.K., N.H., and A.J. wrote the manuscript. All authors discussed the results and commented on the manuscript.

**Competing financial interests**

The authors declare no competing financial interests.

**Data and materials availability**

All (other) data needed to evaluate the conclusions in the paper are present in the paper or the Supplementary Materials. The materials that support the findings of this study are available from the corresponding author upon reasonable and well-intentioned request.

**Supplementary Materials**

Materials and Methods

Figs. S1 to S10
References (29, 30)

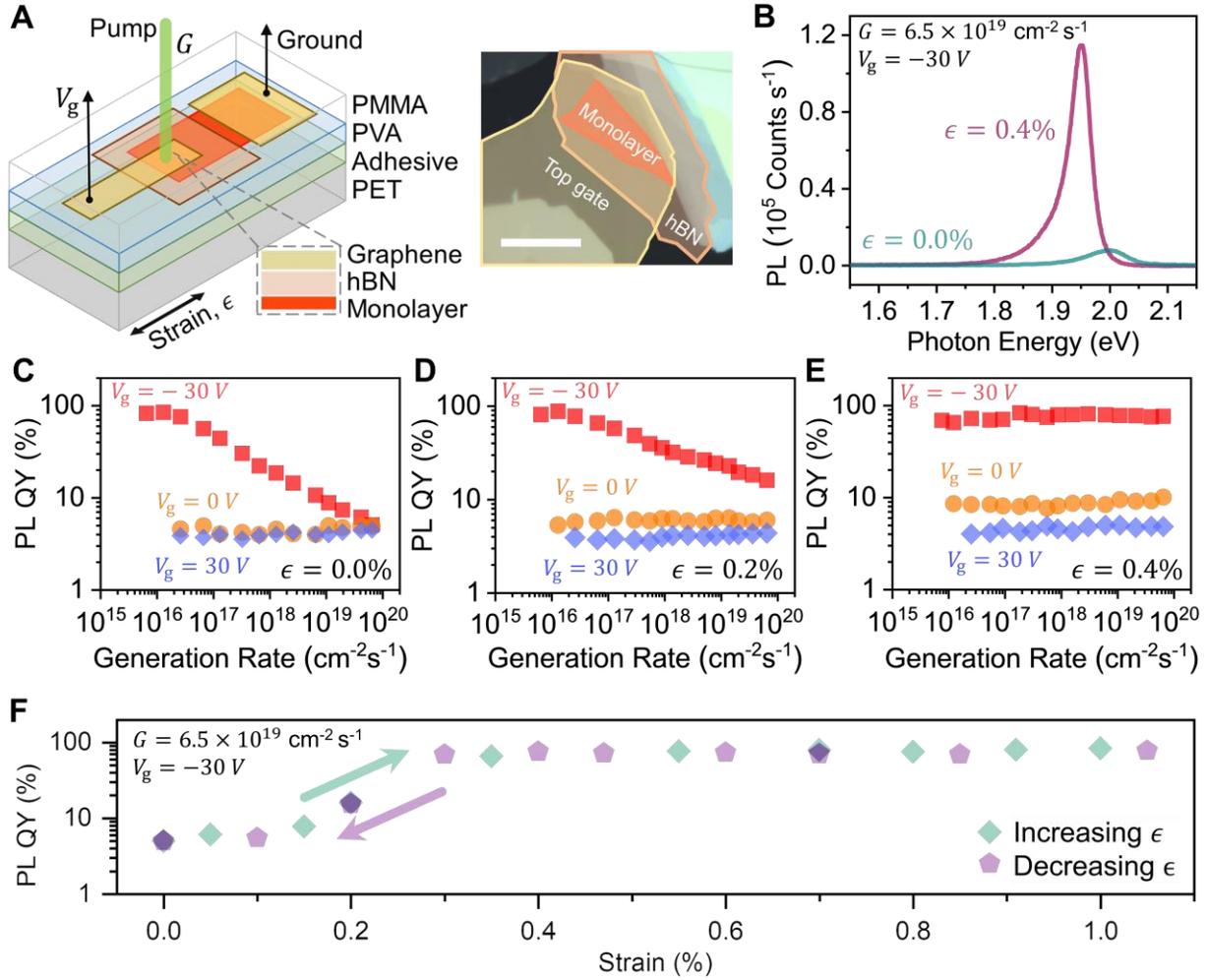

**Fig. 1. Near-unity PL QY in WS$_2$ at all generation rates.** (**A**) Schematic and optical micrograph of the device structure, scale bar is 20 $\mu$m. A two-terminal capacitor structure with graphene as source and gate, and hBN as insulator is fabricated on a flexible polymer substrate. (**B**) Comparison of PL spectra of unstrained and 0.4% strained monolayer WS$_2$ at a high generation rate of $G = 6.5 \times 10^{19}$ cm$^{-2}$s$^{-1}$ and a gate voltage of $V_g = -30\ V$. (**C** to **E**) PL QY of monolayer WS$_2$ as a function of gate voltage, generation rate and strain. (**F**) PL QY approaching unity with the application of strain at a high generation rate of $G = 6.5 \times 10^{19}$ cm$^{-2}$s$^{-1}$.

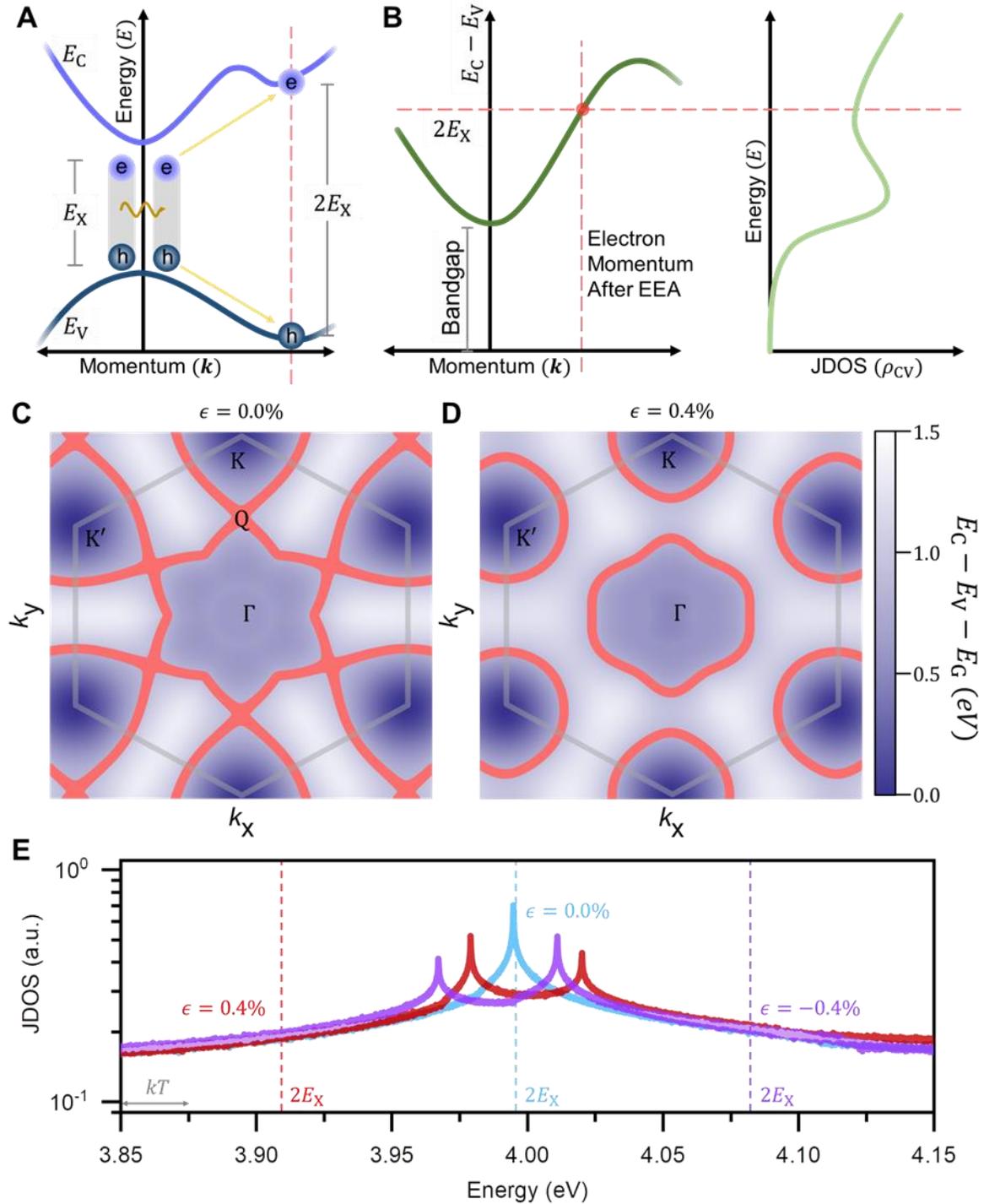

**Fig. 2. EEA suppression by strain. (A)** Schematic describing microscopic mechanism of exciton-exciton annihilation (EEA). EEA occurs when one exciton ionizes another exciton by nonradiatively transferring its energy. **(B)** Momentum and energy conservation dictates that the electron from the ionized exciton ends up with a wavevector where the energy difference between the conduction and valance band is equal to twice the exciton transition energy. The joint density of states (JDOS) at that energy determines the rate of EEA. **(C and D)** Energy difference between the conduction and valance band for unstrained and 0.4% strained monolayer $WS_2$, respectively.

Grey hexagon denotes the first Brillouin zone (each side is 19.946 nm$^{-1}$), whereas red areas are the possible final wavevectors. Possible final wavevectors in unstrained monolayer WS$_2$ include the saddle points Q, where there are VHSs; but strained samples do not. (**E**) Calculated JDOS for monolayer WS$_2$. JDOS $\rho_{\text{CV}}(E)$ at $E = E_{\text{C}} - E_{\text{V}} = 2E_{\text{X}}$ determines EEA rate. Dashed lines denote values of $2E_{\text{X}}$ at the corresponding strain.

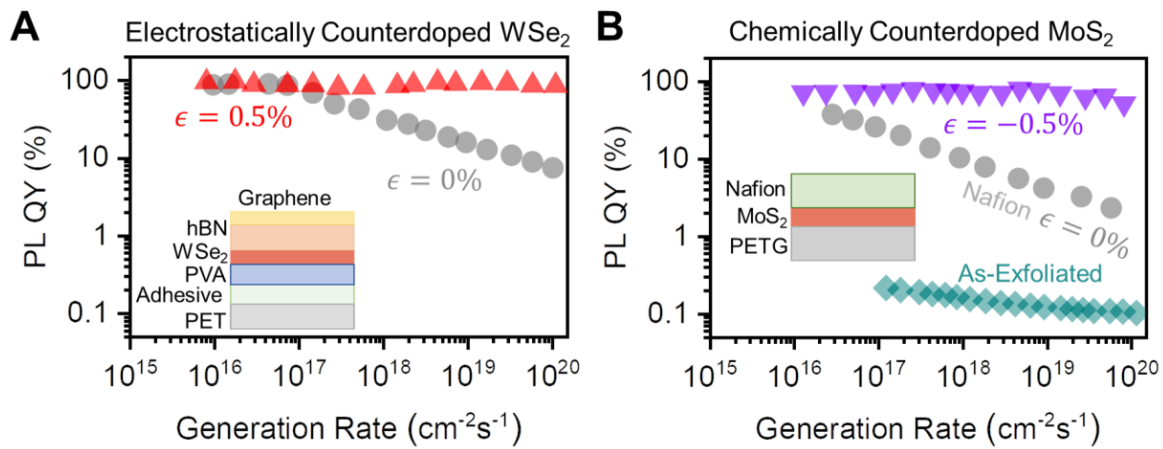

**Fig. 3. General nature of EEA suppression.** Near-unity PL QY at all generation rates in (**A**) electrostatically counterdoped WSe$_2$ and (**B**) chemically counterdoped MoS$_2$ by tensile and compressive strain, respectively, demonstrates the universality of the conditions that suppress nonradiative recombination.

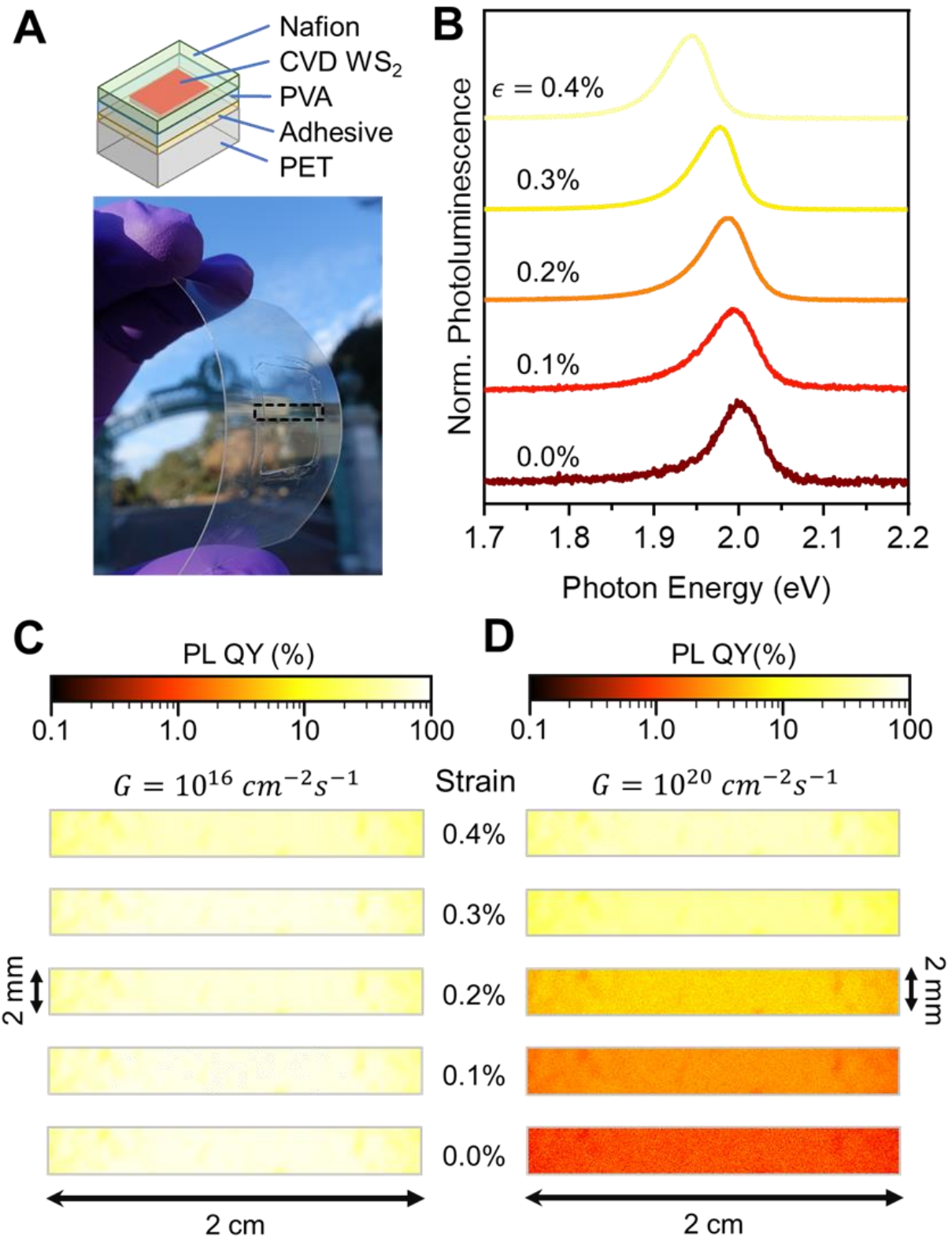

**Fig. 4. | High PL QY on large-area CVD-grown WS$_2$. (A)** Schematic (top) and

photograph (bottom) of the flexible substrate used to strain CVD grown, centimeter-scale WS$_2$.
**(B)** Normalized PL spectra of a typical spot at different strains. (**C** and **D**) Spatial mapping of PL QY of a 2 mm × 2 cm area of grown WS$_2$ at a low generation rate of $G = 10^{16}\ cm^{-2}s^{-1}$ and a high generation rate of $G = 10^{20}\ cm^{-2}s^{-1}$, respectively, shows high PL QY is uniformly achieved by strain at the highest generation rate, while maintaining high PL QY at low pump.

Supplementary Material for

# Inhibited nonradiative decay at all exciton densities in monolayer semiconductors

*Hyungjin Kim[1,2,†], Shiekh Zia Uddin[1,2,†], Naoki Higashitarumizu[1,2], Eran Rabani[2,3,4] and Ali Javey[1,2,*]*

[1]Electrical Engineering and Computer Sciences, University of California, Berkeley, CA 94720, USA
[2]Materials Sciences Division, Lawrence Berkeley National Laboratory, Berkeley, CA 94720, USA
[3]Department of Chemistry, University of California, Berkeley, Berkeley, California 94720, USA
[4]The Raymond and Beverly Sackler Center of Computational Molecular and Materials Science, Tel Aviv University, Tel Aviv 69978, Israel
[†]These authors contributed equally
*Address correspondence to: ajavey@berkeley.edu

**This PDF file includes:**

Materials and Methods
Figs. S1 to S10

References (29, 30)

**MATERIALS AND METHODS**

Device Fabrication
TMDC monolayers (WS$_2$ and WSe$_2$ from HQ Graphene, MoS$_2$ from SPI supplies) were mechanically exfoliated on top of 50 nm SiO$_2$/Si p$^{++}$ substrates and were identified by optical contrast. 10 wt% polyvinyl alcohol (PVA, Alfa Aesar, 98–99% hydrolyzed, molecular weight 130,000 g/mol) solution were spin-coated at 1000 rpm for 40 seconds on the substrate with TMDC monolayer, and baked at 70 °C for 3 minutes to remove water solvent. Subsequently, a polyethylene terephthalate (PET, 125 μm thick) film is attached on top of the PVA film using adhesive (Super glue, Gorilla Glue Company) for ease of handling. Next, the entire flexible substrate (PVA/adhesive/PET) with the encapsulated monolayer TMDC is slowly peeled off from SiO$_2$ substrate using tweezer and resized so that the monolayer is at the center. Hexagonal boron-nitride (hBN, HQ Graphene) is mechanically exfoliated onto a 1×1 cm polydimethylsiloxane (PDMS) substrate. The flexible substrate is put on a sample stage under a modified optical microscope (bh2, Olympus) at room temperature and the hBN-exfoliated PDMS stamp was positioned upside down and aligned with the monolayer on the flexible substrate. The stamp was slowly brought into contact with the monolayer and peeled off rapidly, leading to transferred hBN on top of TMDC monolayer forming the insulator. In a similar fashion, Graphene (Graphenium, NGS Naturgraphit) is subsequently stamp-transferred on top of the TMDC monolayer/hBN stack to from source and top gate. Contacts are formed to the source and gate by applying silver conductive paste (EMS #12640, Electron Microscopy Sciences) to the graphene with a sharpened wooden tip under a Nikon stereomicroscope. Lastly, the device is encapsulated with poly(methyl methacrylate) (PMMA; 950 A11, Microchem) prior to measurement by dropcasting and subsequent soft-baking at 70 °C for 5 minutes.
To prepare compressively strained monolayer MoS$_2$, first MoS$_2$ was exfoliated on PDMS and then stamp was slowly brought into contact with the center of a piece of glycol-modified Polyethylene terephthalate (PETG, 2.54 mm thick, 3 cm length) at 90° C and peeled off, followed by rapid quenching of the MoS$_2$-transferred PETG down to room temperature. The coefficients of thermal expansion (CTEs) of PETG and monolayer MoS$_2$ over the range of 20-90 °C were experimentally measured to be 80.5 ± 2.5 ppm/°C and 7 ppm/°C, respectively (*29*). The biaxial compressive strain applied in the monolayer MoS$_2$ due to the CTE mismatch can thus be calculated as

$$\varepsilon(T_{tr}) = \int_{20°C}^{T_{tr}} \alpha_{PETG}(T)dT - \int_{20°C}^{T_{tr}} \alpha_{MoS_2}(T)dT \qquad (1)$$

where $T_{tr}$ is the transfer temperature of monolayer onto the PETG, $\alpha_{PETG}$ is the CTE of the PETG substrate, and $\alpha_{MoS_2}$ is the CTE of the monolayer MoS$_2$. Since we are using the transfer temperature of 90 °C, the biaxial compressive strain in the monolayer MoS$_2$ resulting from this quenching process was 0.51 %.

After the transfer, 3% Nafion perfluorinated resin solution (5 wt % in lower aliphatic alcohols and water, contains 15−20% water, Sigma Aldrich) diluted by ethanol was spin coated at 4000 rpm on the prepared MoS$_2$ sample with biaxial compressive strain.

Electrical and Optical Characterization
Devices were charged from an Keithley 2410 Source Meter Unit connected to the gate electrode, while the graphene source contact was grounded. The PL QY was measured using a customized micro-PL instrument which was calibrated as following procedure. The PL QY measurement calibration is also described in detail in previous study (*1*, *4*). The wavelength of spectrometer was verified with Ar lamps (Newport) and the wavelength-dependent instrument response function was measured by a virtual Lambertian blackbody source, which was created with a stabilized lamp (Thorlabs SLS201) and a spectralon reflection standard (Labsphere). The collection efficiency was then acquired by measuring the laser response which was focused on the spectralon reflection standard. Because the emission profile of the monolayer semiconductors has been known to be Lambertian, this gives the reasonable approximation of the source of PL emission from the monolayers. For calibration measurements taken from the spectralon reflection standard, the fraction of generated photons which escape from the sample was calculated using $1/4n^2$, where $n$ is the refractive index of the medium (*30*). Independent approach was also used to verify the system calibration using a sample which has been known to have QY ~100% (rhodamine 6G in methanol). Generation rate $G$ is the number of excitons created or the number of photons absorbed per-unit area per unit time.

$$G = \frac{\alpha P}{A\hbar\omega}$$

where, $\alpha$ is the monolayer absorption at pump wavelength, $P$ is the pump laser power in Watts, $A$ is the area of the laser spot in cm$^2$, and $\hbar\omega$ is the energy of a 514.5 nm single photon. The absorption at pump wavelength is independent of strain. We vary the generation rate by varying the incident pump laser power.

In order to ensure that there is no change in light outcoupling factors or collection efficiency, we measured the samples that have PL QY independent of strain, as shown in the fig. S10. All of the samples show no change in PL spectra with applied strains, indicating that substrate-bending does not affect the PL QY and the same calibration approach can be used due to no change in light outcoupling and collection efficiency of the system.

An Ar ion laser with a 514.5 nm line was used as the excitation source. All measurements reported in this paper are taken at room temperature, in an ambient lab conditions under nitrogen flow. The applied strain on the flexible substrate were calculated using the equation $\varepsilon = t/R$, where $2t$ and $R$ are the substrate thickness and curvature radius. The thickness $2t$ of the entire flexible substrate is measured through the cross-section optical image. The curvature radius $R$ were measured through the side-view photograph. All theory calculations are carried out in MATLAB.

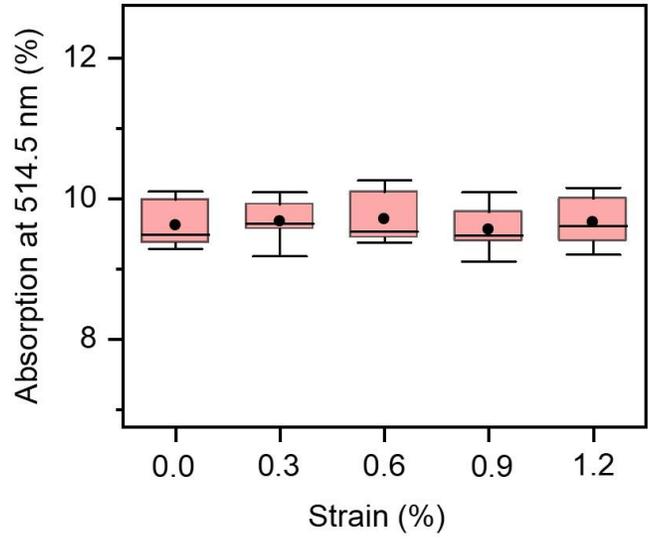

**fig. S1. Absorption at excitation wavelength under applied strain.** Absorption of monolayer $WS_2$ as a function of applied strain at laser wavelength ($\lambda$ = 514.5 nm).

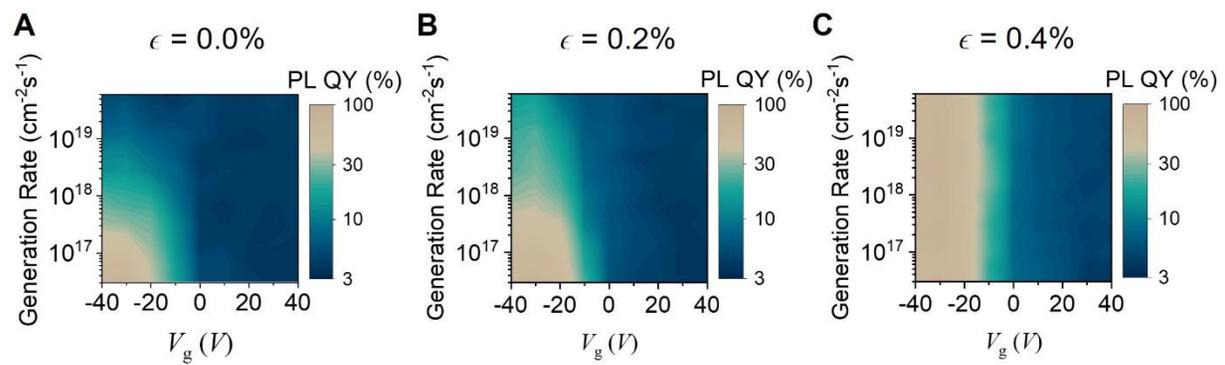

**fig. S2. PL QY versus *G* and *V*$_g$ in WS$_2$ at different strains.** A color plot of PL QY in monolayer WS$_2$ as a function of gate voltages and generation rate with applied strains.

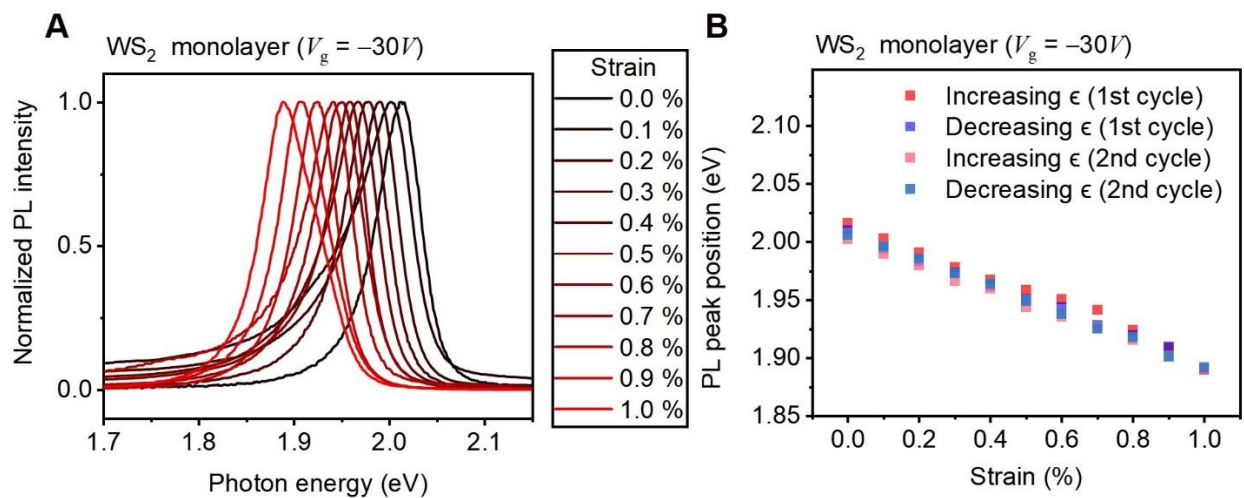

**fig. S3 PL peak position shift of monolayer WS$_2$.** (**A**), Normalized PL intensity of monolayer WS$_2$ at $V_g = -30V$ as a function of applied strains. (**B**), PL peak position during multiple bending and release cycles.

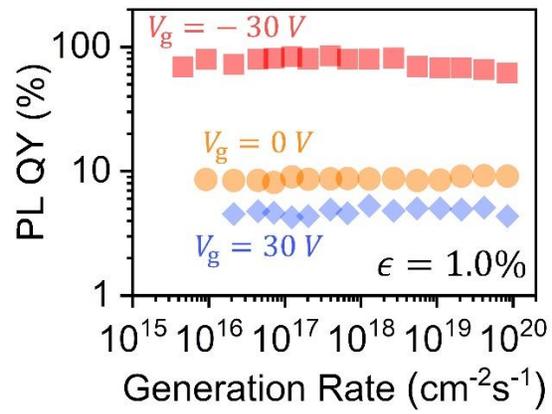

**fig. S4 Near-unity PL QY in WS$_2$ at all generation rates.** PL QY of monolayer WS$_2$ as a function of gate voltage and generation rate at 1.0 % strain.

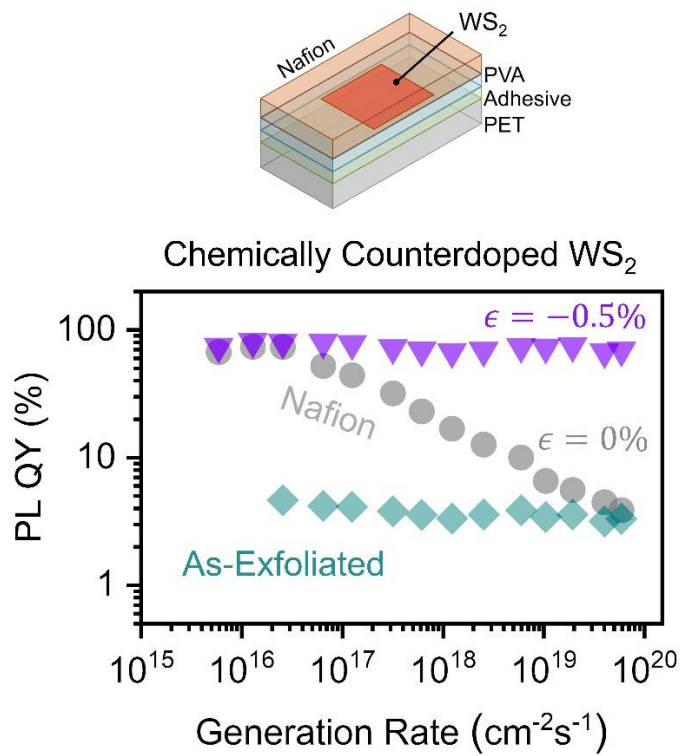

**fig. S5. Near-unity PL QY in chemically counterdoped WS₂ under strain.** PL QY of as-exfoliated and chemically counterdoped monolayer WS$_2$ as a function of strain.

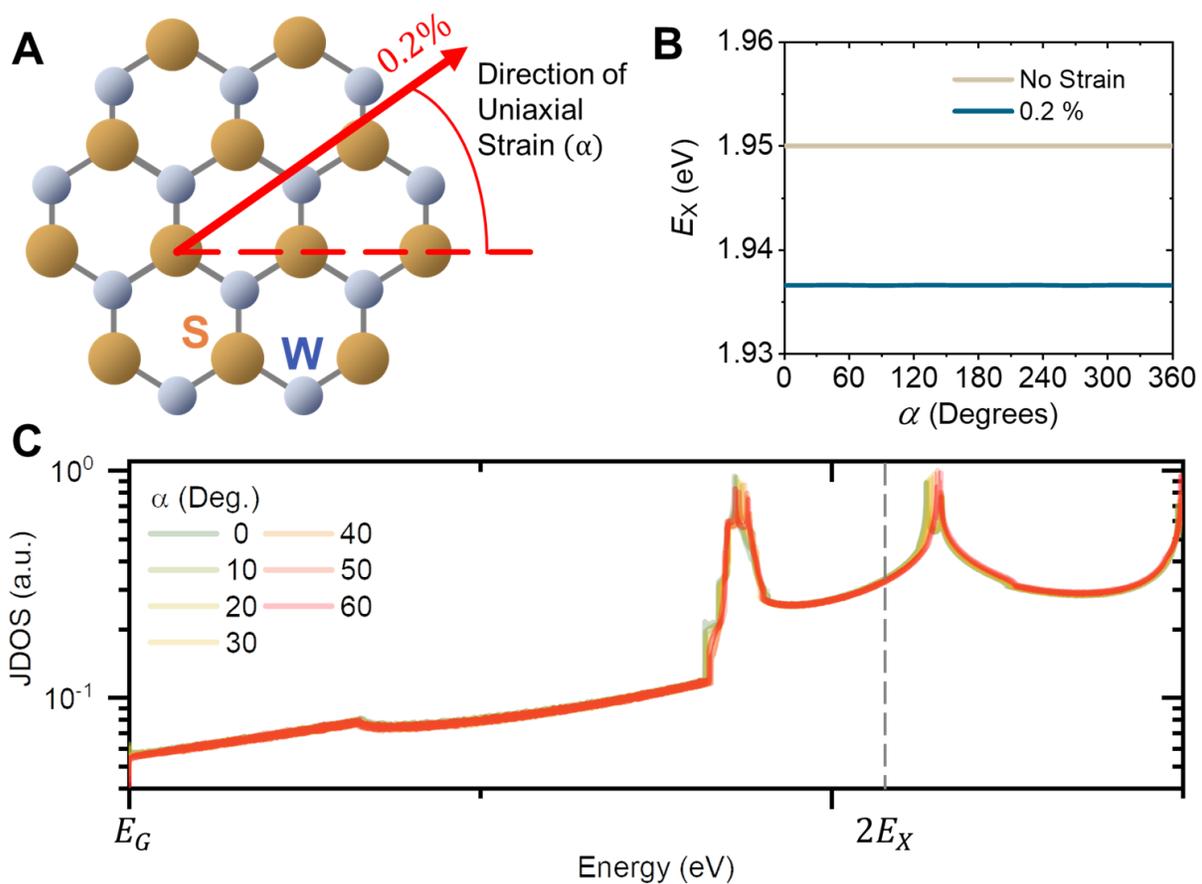

**fig. S6. Direction independence of EEA suppression.** (**A**), Direction of uniaxial strain (**B**), Exciton transition energy as a function of strain direction (**C**), VHS is shifted from twice of exciton transition energy independent of tensile strain direction.

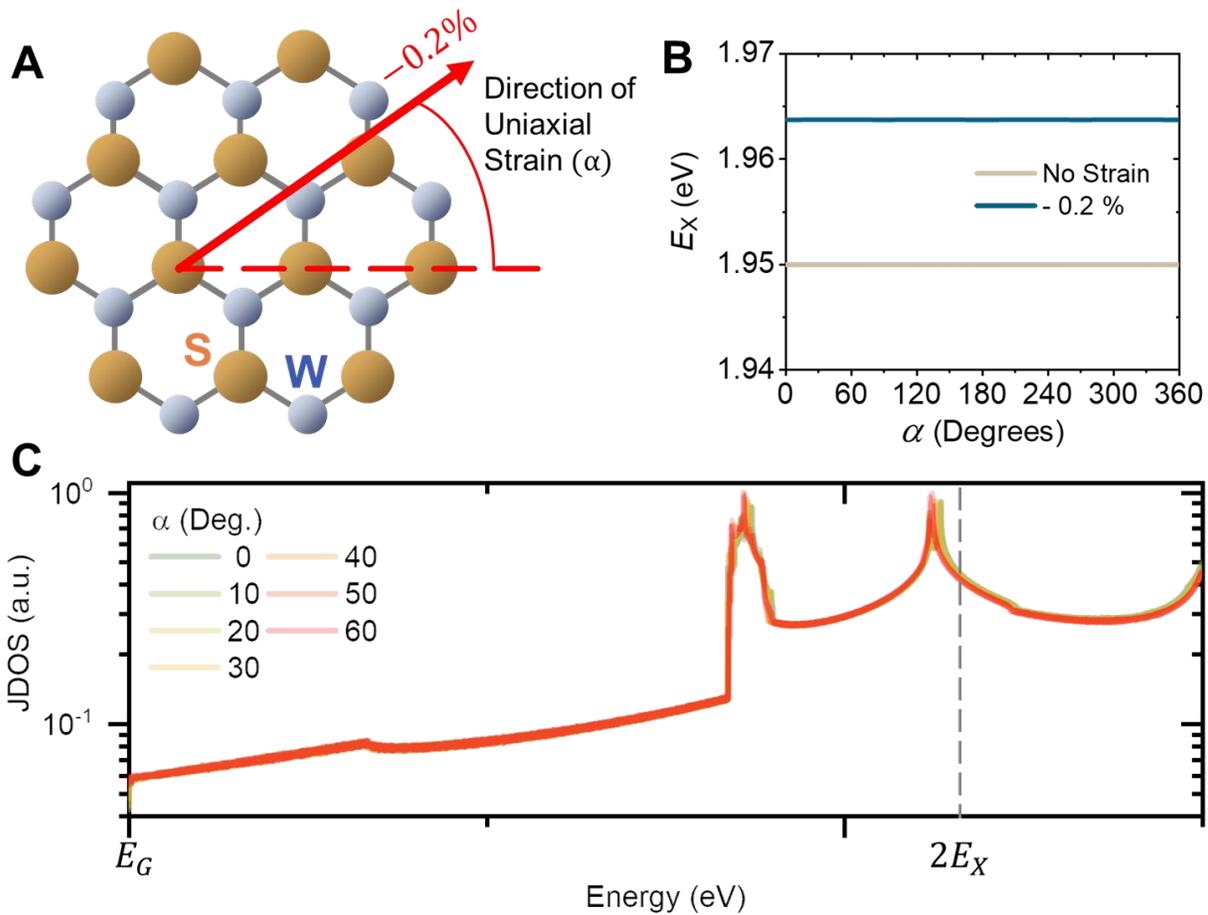

**fig. S7. Direction independence of EEA suppression.** (**A**), Direction of uniaxial strain (**B**), Exciton transition energy as a function of strain direction (**C**), VHS is shifted from twice of exciton transition energy independent of compressive strain direction.

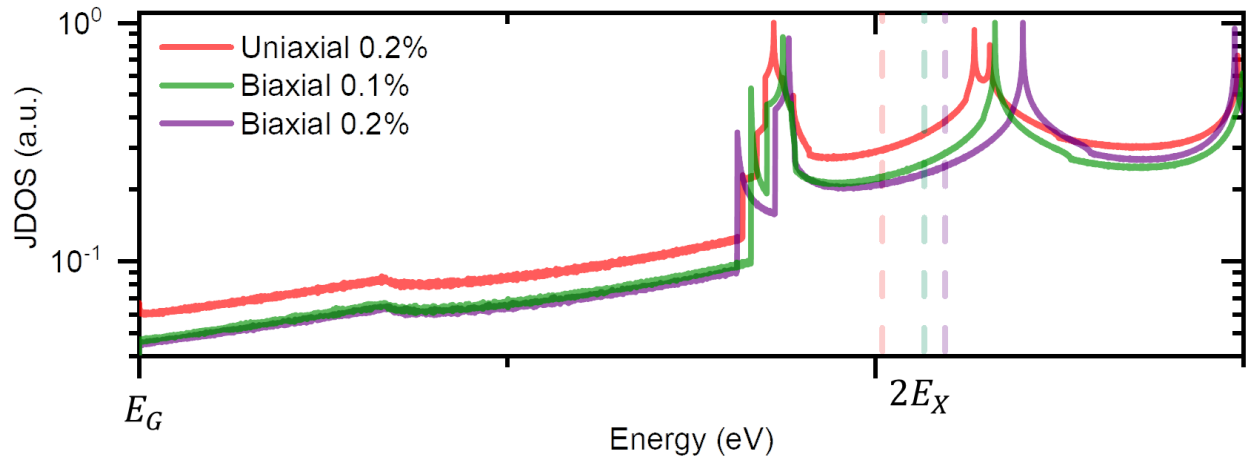

**fig. S8. EEA Suppression with Biaxial Strain.** VHS is also shifted from twice of exciton transition energy with biaxial compressive strain.

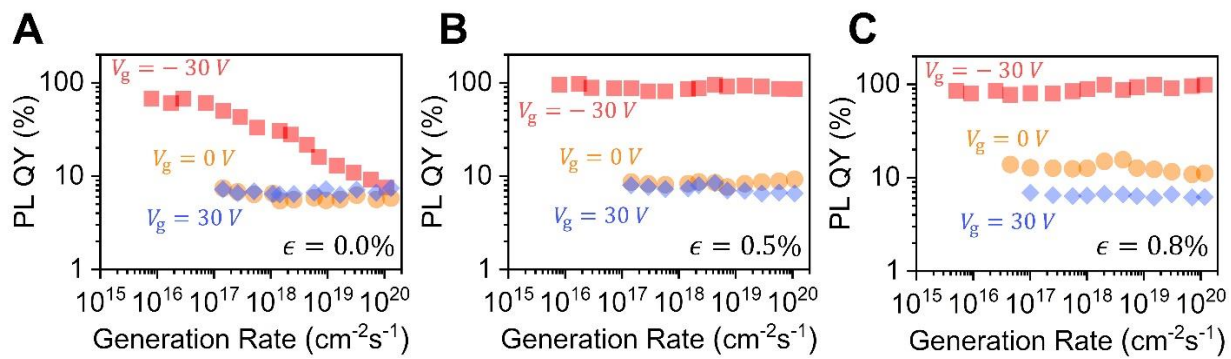

**fig. S9. Near-unity PL QY in WSe$_2$ at all generation rates.** (**A**), (**B**), (**C**), PL QY of monolayer WSe$_2$ as a function of gate voltage, generation rate and strain.

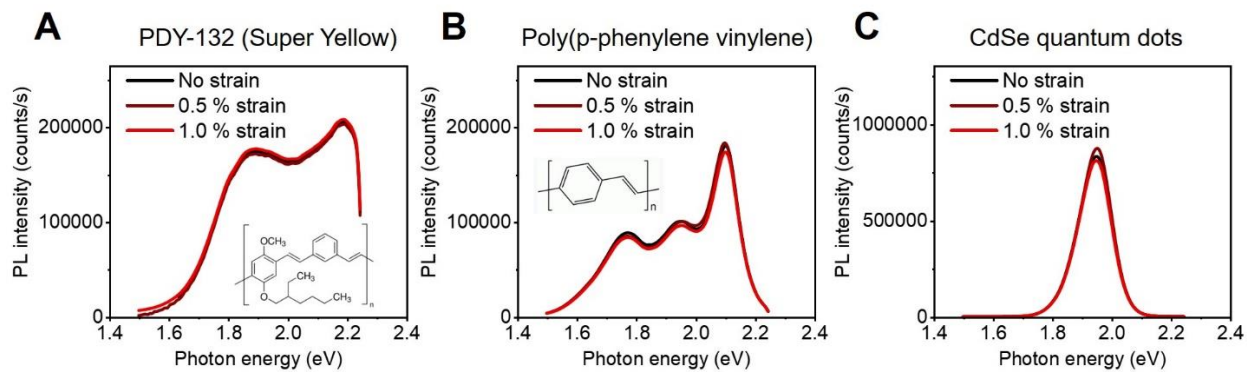

**fig. S10. Negligible change in PL by bending of substrates.** PL spectra of (**A**) PDY-132, (**B**) Poly(p-phenylene vinylene) (PPV), (**C**) CdSe quantum dots on PVA-PET flexible substrates as a function of applied strains.